# On the modification of the photoionization of subvalent subshells in endohedral noble gas atoms

M. Ya. Amusia[1, 2], A. S. Baltenkov[3, 4], A. Z. Msezane[5], and S. T. Manson[3]

[1]Racah Institute of Physics, The Hebrew University, Jerusalem 91904, Israel
[2]Ioffe Physical-Technical Institute, St.-Petersburg 194021, Russia
[3]Department of Physics and Astronomy, Georgia State University, Atlanta, GA 30303, USA
[4]Arifov Institute of Electronics, Tashkent, 700125, Uzbekistan
[5]Center for Theoretical Studies of Physical Systems, Clark Atlanta University, Atlanta, GA 30314, USA

**Abstract**

It is demonstrated that outer and inner electron shells, including that formed by collectivized electrons of the fullerene $C_{60}$, affects dramatically the cross section of the subvalent $ns$-subshells of the noble gas endohedral atoms $A@C_{60}$. The calculations are performed within the framework of a very simple, so-called "orange skin", model that makes it possible, in spite of its simplicity, to take into account the modification of the $ns$-subshell due to its interaction with inner and outer atomic shells, as well as with the collectivized electrons of the $C_{60}$. As a concrete example, we consider the Xe $5s$ electrons completely collectivized by the powerful action of the Xe close and remote multi-electron neighboring shells.

## 1. Introduction

The pronounced action of the multi-electron neighboring $5p^6$ and $4d^{10}$ subshells upon a few electron one, considering as an example the $5s^2$ subshell in Xe, was predicted in the frame of so-called Random Phase Approximation with Exchange (RPAE) approximately thirty years ago [1] and soon confirmed experimentally [2]. Since that time the photoionization of $5s^2$ in Xe and its neighbors like I, Cs and Ba, has become a subject of a number of investigations [3, 4].

The physical nature of these so-called intershell effects is as follows. A many-electron atomic subshell is polarized, or in other words, virtually excited by an incoming electromagnetic wave and therefore a time-dependent dipole moment is induced. Under the action of this dipole moment a neighboring atomic subshell is ionized. RPAE is extremely convenient to describe this effect. Since the electronic subshells in an atom are not spatially separated well enough, the amplitude of this two-step photo-process cannot be expressed accurately enough directly via the dipole polarizabilities of the many-electron subshells.

In this sense the situation for the endohedral atoms $A@C_{60}$ is quite different. The radius of fullerene $C_{60}$ shell significantly exceeds that of an encapsulated atom. This makes it possible for photoionization of the A atom, in the first approximation, to consider the electronic sub-systems of the fullerene shell and atom as practically independent of each other. For this reason, the amplitude of atom photoionization going through virtual excitation of the $C_{60}$ shell electrons can be expressed directly via the dynamic polarizability of the fullerene shell $\alpha_{C_{60}}(\omega)$. In those cases when the frequency of electromagnetic radiation is close to frequencies of plasma oscillations of the collectivized electrons of the $C_{60}$ shell, a role of this two-step process becomes decisively important, as the role of $4d^{10}$ upon $5s^2$ in isolated Xe.

It has been demonstrated recently that the photoionization cross section of the Xe 5*s* subshell in Xe@$C_{60}$ is strongly modified due to reflection of the photoelectron's wave by the fullerene shell [5]. A simple method was developed to take into account this process [6]. The potential of the $C_{60}$ shell was presented in [6] by of a zero-thickness $\delta$-type bubble potential. In this approach the inner degrees of freedom of the $C_{60}$ shell, namely its ability to be polarized, were neglected.

However, it is known from experiment [7] that $C_{60}$ has a dipole polarizability that is extremely big in atomic scale. The frequency dependence of $\alpha_{C_{60}}(\omega)$ exhibits an almost symmetrical powerful maximum at the so-called Giant resonance frequency that is of about 20 eV. This energy is close to the ionization potential of the Xe 5*s* subshell. Having in mind the strong modifications of the photoionization cross sections under the action of intershell interaction in isolated atoms [1], we can expect strong intershell effects for endohedral atoms as well.

Indeed, as we will demonstrate in this paper, the photoionization of Xe 5*s* subshell in the endohedral system Xe@$C_{60}$ is a remarkable concrete example illustrating the role of the intershell interactions in the fullerene-like molecules, qualitatively similar but even stronger than in the isolated atoms.

The aim of this paper is to clarify how the dynamic response of the multi-electron $C_{60}$ shell affects the photoionization of the electron subshells of the endohedral atom. The method developed can be applied to other objects where many other atoms with collectivized electrons surround a given atom or an ion.

## 2. Essential formulae

The photoionization characteristics of very many complex atoms were calculated within the framework of RPAE that takes into account along with the direct ionization amplitude of the considered electrons $d_s$ (in our case, these are the 5*s* electrons of Xe) the dipole polarization of other electron shells. The polarized shells ionize the *s*-electron due to inter-shell interaction. The approach developed for isolated atoms can be equally well applied to systems with other electron shells like endohedral atoms.

Symbolically, the total amplitude of some *s*-electron ionization $D_s$ can be presented as a sum of two terms [8]

$$\hat{D}_s = \hat{d}_s + \hat{D}_o \hat{\chi} U_{os}, \qquad (1)$$

where $\hat{D}_o$ is the ionization amplitude of any other than "*s*"-electrons, $\hat{\chi} = 1/(\omega - \hat{H}_{ev}) - 1/(\omega + \hat{H}_{ev})$ is the propagator of other electron excitation, i.e. electron-vacancy pair creation, $\hat{H}_{ev}$ is the pair Hartree-Fock Hamiltonian and $U_{os} \equiv V_{os,dir} - V_{os,exch}$, with $V_{os,dir}$ and $V_{os,exch}$ being the operator of direct and exchange pure Coulomb interaction between "o" and "s" electrons.

We concentrate on almost spherically symmetric systems. The formula (1) is simplified considerably if the "o"-electrons are either at much smaller or much large distance from the center of the system than the "s"-electrons. In both cases the Coulomb interaction is considerably simplified, becoming either

$$U_{os} \approx \mathbf{r}_s \cdot \mathbf{r}_o / r_s^3, \text{ (for } r_s \gg r_o), \tag{2a}$$

or

$$U_{os} \approx \mathbf{r}_s \cdot \mathbf{r}_o / r_o^3, \text{ (for } r_o \gg r_s). \tag{2b}$$

Here $\mathbf{r}_s$ and $\mathbf{r}_o$ are the "s"- and "o"-electron shells radii, respectively.

In the language of many-body diagrams the expression (1) can be presented as

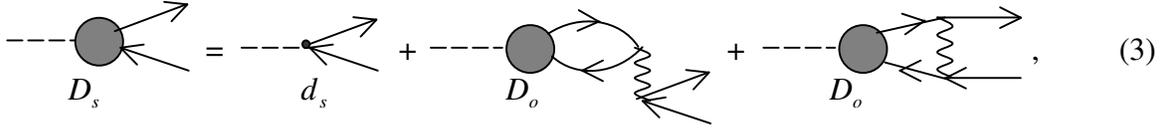

$$\tag{3}$$

where the dashed line, line with arrow to the right (left) and the wavy line represent the incoming photon, electron, vacancy and the direct (exchange) Coulomb "o-s" electrons interaction, respectively.

The equation (1) and (3) can be easily generalized in the spirit of the Landau Fermi-liquid theory by incorporating into $d_s$ all but simple electron-vacancy excitations, for example, two electron – two vacancy excitations of the s-shell [9]

The effect of the "o"-shell is represented particularly simple, when it is an inner one, located well inside the intermediate and outer atomic subshell. Then rightfully neglecting the exchange "o-s"-interaction and representing $U_{os}$ as (2a), one reduces (1) to an algebraic equation instead of operator one where $\hat{D}_o \hat{\chi} U_{os}$ is substituted by the following expression

$$[2 \sum_{evexit,o} \omega_{ev} D_{ev}(\omega)(\omega^2 - \omega_{ev}^2)^{-1} d_{ev}] / r_s^3 \equiv -\alpha_o(\omega) / r_s^3, \tag{4}$$

where $\alpha_o(\omega)$ is the total dipole polarizability of the "o"-shell. The summation in (4) *evexit,o* goes over all electron-vacancy excitation of the considered shell. It is thus assumed that more complex excitations are included, in the spirit of Landau Fermi-liquid theory into $d_{ev}$ and via (1) into the amplitude $D_{ev}(\omega)$ as well.

In (4) we used an alternative definition of $\alpha_o(\omega)$. Usually, it is defined as

$$\alpha_o(\omega) \equiv -[2 \sum_{evexit,o} \omega_{ev} | D_{ev}(\omega_{ev}) |^2 (\omega^2 - \omega_{ev}^2)^{-1}], \tag{5}$$

but it can be easily demonstrated that this definition and that in (4) are identical [9, 10].

Thus, in the case of an inner shell "o" one has instead of Eq. (1) the following formula [10]:

$$D_s(\omega) \cong d_s \left(1 - \frac{\alpha_o(\omega)}{r_s^3}\right), \tag{6}$$

Usually, of interest is $D_s(\omega)$ for photon energy $\omega^*$ of an order of $I_s$ – the $s$-electron ionization potential. The outer shell photoionization cross section has its highest values at $\omega \sim I_s$. Since $I_s \ll I_o$, one can substitute $\omega$ in $\alpha_o(\omega)$ of (6) by zero, having instead of dynamic $\alpha_o(\omega)$ the static dipole polarizability $\alpha_o$.

If one considers as "o" the outer electrons, with $r_o \gg r_s$, the expression Eq. (1) is again considerably simplified. The exchange can be again neglected. Usually, the "o"-shell is a layer of electrons the thickness of which is much smaller than its radius $r_o$. In this case one obtains from consideration performed in [10] an expression similar to Eq. (6) where $r_s^3$ is substituted by a mean value $\bar{r}_o^3$:

$$D_s(\omega) \cong d_s\left(1 - \frac{\alpha_o(\omega)}{\bar{r}_o^3}\right). \qquad (7)$$

For $\omega$ above the $s$-subshell ionization threshold, since for $\omega > I_s \gg I_o$ the polarizability $\alpha_o(\omega)$ from Eq. (4) can be simplified by neglecting $\omega_{ev}$ as compared to $\omega$, and thus substituted by its dynamic high $\omega$ limit, $\alpha_o(\omega) = -N_o/\omega^2$. This correction for an isolated atom is small since $\omega^2 \sim I_s^2 \sim N_s/r_s^3$, that means substituting $\alpha_o(\omega)/\bar{r}_o^3$ by $N_o r_s^3 / N_s r_o^3 \ll 1$.

For isolated atoms Eq. (1) is usually solved numerically, without additional simplifications connected to the smallness of the ratio $r_s/r_o$ or $r_o/r_s$. Then the effect of any additional electrons, e. g. those belonging to the $C_{60}$ shell in endohedral atoms can be treated as something on top of the atomic multi-electron effects. An essential simplification comes from the fact, that the $C_{60}$ radius $R_c$ is big enough, $R_c \gg r_s$. It is also essential that the electrons in $C_{60}$ are located within a layer the thickness of which $\Delta R_c$ is considerably smaller than $R_c$. In this case, one comes to an expression similar to Eq. (6), except as $d_s$ is substituted by $D_s^{(A)}(\omega)$, i.e. by the amplitude of $s$-electron photoionization with all essential atomic correlations taken into account:

$$D_s(\omega) \cong D_s^{(A)}(\omega)\left(1 - \frac{\alpha_o(\omega)}{R_c^3}\right). \qquad (8)$$

Note that a similar expression has been suggested recently in Ref. [12] on the basis of pure classical consideration. In our derivation, just as in Ref. [12], we have neglected the higher order corrections that can be estimated as powers of $\alpha^{(A)}(\omega)/R_c^3$. The latter parameter is, except discrete level resonances, much smaller than one.

Using the relation between the imaginary part of the polarizability and the photoabsorption cross-section, namely $\mathrm{Im}\,\alpha_{C_{60}}(\omega) = c\sigma_{C_{60}}(\omega)/4\pi\omega$, one can derive the polarizability of the C60 shell. Although experiments [7, 12, 13] provide no direct absolute values of $\sigma_{C_{60}}(\omega)$, it can be reliably estimated using different normalization procedures based of the sum rule: $(c/2\pi^2)\int_{I_o}^{\infty}\sigma_{C_{60}}(\omega)d\omega = N$, where $N$ is the number of collectivized electrons. The dispersion relation

$$\text{Re}\,\alpha_{C_{60}}(\omega) = \frac{c}{2\pi^2}\int_{I_{60}}^{\infty}\frac{\sigma_{C_{60}}(\omega')d\omega'}{\omega'^2 - \omega^2} \qquad (9)$$

connects the real part of the polarizability $\text{Re}\,\alpha_{C_{60}}(\omega)$ with the imaginary part $\text{Im}\,\alpha_{C_{60}}(\omega)$. In (9) $I_{60}$ is the $C_{60}$ ionization potential.

---

* The atomic system of units: $e = m = \hbar = 1$ is used throughout this paper

This approach was used for deriving the polarizability of $C_{60}$ in [14], where it was considered that $N = 240$, i.e. 4 electrons collectivized from each C atom. Using the photoabsorption data that are considered in [12] as most reliable, we obtained $N_{eff} = 255$. This is sufficiently close to the value, accepted in [14]. Note that since the one-electron photoionization cross-section of $C_{60}^+$ is much smaller than similar cross section for $C_{60}$, one cannot limit with this cross section, measured and calculated in [13]. Obviously, other photoionization channels are of importance and have to be included.

As seen from the experimental photoabsorption cross section, it is small at threshold (which also means relatively low intensity of discrete excitations) and is dominated by a huge maximum well above the threshold. Therefore, in Eq. (9) a small contribution of discrete excitations of the $C_{60}$ collectivized electrons are neglected.

In order to take into account the processes of reflection and refraction of the photoelectron's wave by the fullerene shell, we use a $\delta$- bubble model that represents the static $C_{60}$ potential as $U(r) = -V_0\delta(r - R_c)$, with $V_0$ chosen in such a way as to reproduce the binding energy of $C_{60}^-$ negative ion. In the frame of this model potential, its influence upon the photoionization amplitude is presented by a factor $F_A(\omega)$, which is of oscillatory nature and takes into account reflection of the p-photoelectron by the $C_{60}$ shell. The details of calculation of $F_A(\omega)$ can be found in Ref. [5]. In short, this function is expressed via the regular and irregular at $r = 0$ photoelectron wave functions.

Entirely, the following relation gives the amplitude $D_s(\omega)$ for the Xe atom inside the $C_{60}$ shell:

$$D_s(\omega) = F_A(\omega)D_s^{(A)}(\omega)\left(1 - \frac{\alpha_{C_{60}}(\omega)}{R_{C_{60}}^3}\right). \qquad (10)$$

Using this amplitude, one has for the cross section

$$\sigma_s(\omega) = \sigma_s^{(A)}(\omega)|F_A(\omega)|^2\left|1 - \frac{\alpha_{C_{60}}(\omega)}{R_{C_{60}}^3}\right|^2 \approx \sigma_s^{(A)}(\omega)|F_A(\omega)|^2\left[1 + \frac{\text{Im}^2\alpha_{C_{60}}(\omega)}{R_{C_{60}}^3}\right]. \qquad (11)$$

The latter approximate equality is valid over the range near the maximum of the $C_{60}$ Giant resonance where $\text{Im}\,\alpha_{C_{60}}(\omega)$ is considerably bigger than $\text{Re}\,\alpha_{C_{60}}(\omega)$.

## 3. Results of calculations

The results of calculations for the function $G_{C_{60}}(\omega) = |1 - \alpha_{C_{60}}(\omega)/R_{C_{60}}^3|^2$ are presented in Fig.1.

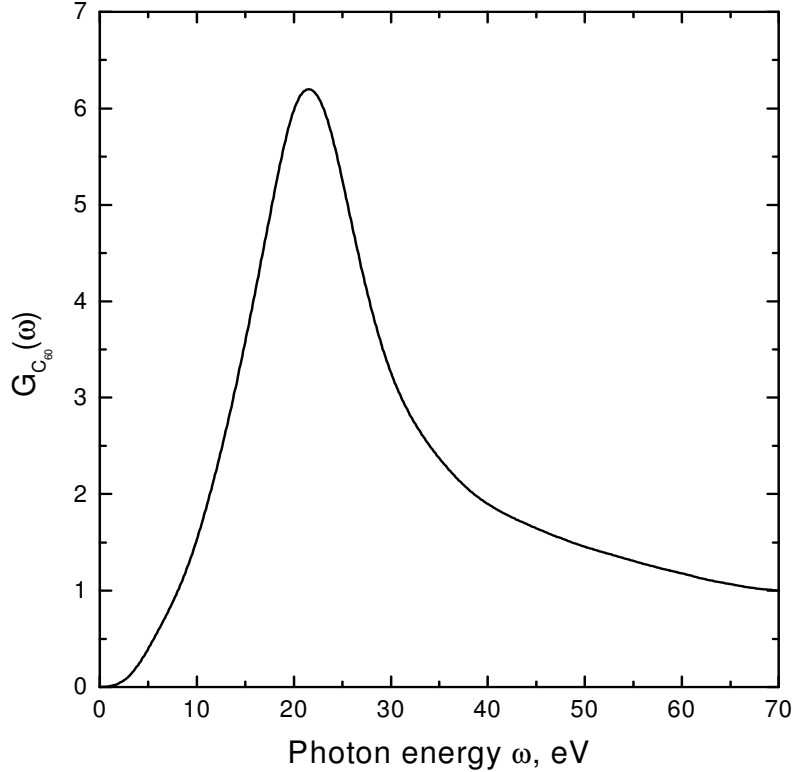

Fig.1 Factor $G_{C_{60}}$ as a function of photon energy $\omega$.

It is seen that it has a powerful maximum that reaches a sufficiently prominent value of about $G_{C_{60}}(\omega) \approx 6.25$ at the Giant resonance frequency $\omega \approx 20$ eV. It should be noted at the same time that at $\omega = 0$ the factor $G_{C_{60}}(\omega)$ is equal to zero. It seems that this is a general feature of this function, since a static electric field cannot penetrate through the conducted $C_{60}$ shell.

Already at $\omega = 70$ eV the factor $G_{C_{60}}(\omega)$ is close to unity. It is essential to have in mind that, as it is evident from Fig. 1, the effects of dynamic polarizability of the fullerene are pronounced up to 60 eV. In general, it is seen that the dynamic reaction of the fullerene shell enhances the effects of reflection of photoelectrons from Xe 5s-subshell by the $C_{60}$ shell.

Fig. 2 presents the results for the photoionization cross-section of the 5s electrons in Xe@$C_{60}$. It is seen that the common efforts of static and dynamic reaction of the $C_{60}$ shell leave almost nothing similar to the pure atomic cross section. The confinement resonance at about 27 eV is strongly enhanced while resonance modification at 50 eV is only about 30%. Pronounced are the effects of the $C_{60}$ shell even up to 100 eV.

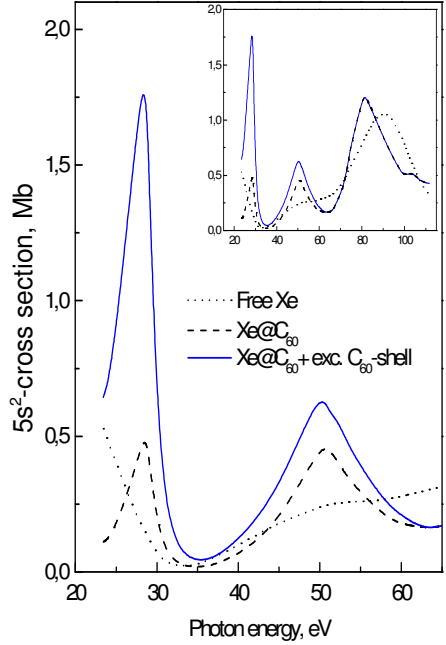

Fig. 2. Photoionization cross-section of the 5*s* electrons in Xe@$C_{60}$. Dotted line is the isolated atom, dashed line - Xe@$C_{60}$, with account of the static $C_{60}$ potential. Solid line: the same, with taking into account of $C_{60}$ dynamical polarizability.

**4. Discussion**

It was demonstrated in this paper that dynamic reaction of the $C_{60}$ collectivized electrons modifies essentially the photoionization cross section of the inner atom in A@$C_{60}$ at relatively low energy of incoming photons. Just as in pure classical approach [11], the here-performed quantum mechanical calculations demonstrate that the inner atom at $\omega < 5$ eV is totally screened while starting from $\omega > 10$ eV and up to $\omega < 80$ eV the cross section becomes enhanced.

The angular distributions of the photoelectrons and their spin polarization have to be also modified by the dynamic reaction of the $C_{60}$ shell. In this case for non-dipole terms of these parameters the quadrupole collective excitations of $C_{60}$ could be of importance. Unfortunately, almost nothing is known about these excitations until now.

It should be noted in conclusion that, as it follows from the above-given consideration, the effect of dynamical reaction of the electronic sub-system of the fullerene shell on the process of photoionization of encapsulated atoms is completely defined by a universal function $G_{C_{60}}(\omega)$ being the same for any endohedral systems A@$C_{60}$. This fact significantly extends a possibility of using the bubble model to describe the processes of electromagnetic radiation interaction with fullerene-like molecules.

## 5. Acknowledgements

The authors are grateful for financial support to Bi-national Science Foundation, Grant 2002064, Israeli Science Foundation, Grant 174/03. M.Ya. Amusia and A. S. Baltenkov are grateful to CTPS at CAU and GSU (Atlanta, GA) for hospitality. This work was also supported by Uzbekistan National Foundation, Grant Ф-2-1-12.